\documentclass[aps,prb,twocolumn,showpacs,superscriptaddress]{revtex4}
\usepackage{graphicx}
\usepackage{amsmath}
\usepackage{amssymb}
\usepackage{pifont}
\usepackage{color}

\begin{document}

\title{On the Green-Kubo relationship of the liquid-solid friction coefficient}
\author{Lyd\'eric Bocquet}
\affiliation{Institut Lumi\`ere Mati\`ere, Universit\'e Lyon 1 - CNRS, UMR 5306,
 Université de Lyon, 69622 Villeurbanne cedex, France}
\author{Jean-Louis Barrat}
\affiliation{LiPhy, Universit\'e Joseph Fourier - CNRS, UMR 5588,
 38402 Grenoble Cedex, France}

\begin{abstract}
In this paper, we propose a new derivation for the Green-Kubo relationship for the liquid-solid friction coefficient,
characterizing hydrodynamic slippage at a wall.
It is based on a general Langevin approach for the fluctuating wall velocity, involving a non-markovian memory kernel with vanishing
time integral. The calculation highlights some subtleties of the wall-liquid dynamics, leading to superdiffusive motion of the
fluctuating wall position.
\end{abstract}
\date{\today}
%nopacs
\maketitle

\section{Introduction}

The question of the hydrodynamic boundary conditions applicable at a solid-liquid interface has raised  considerable interest over the last years \cite{LaugaBrenner,NetoCraig,Bocquet2007,Bocquet2010} . A fundamental understanding of dynamics of fluids at interfaces is now emerging, due to an intense theoretical and numerical work on the topics, see {\it e.g.} \cite{Bocquet2007} for a review, as well as the development of new experimental tools allowing to investigate fluid dynamics at the nanoscale \cite{Bocquet2010}. 
Deviations from the no-slip boundary conditions are expected to impact drastically fluid transport in micro- and nano- channels by allowing slippage at its boundaries and reducing accordingly the dissipation \cite{Holt06}. From a fundamental perspective, slippage is usually accounted for by the so-called Navier partial slip boundary condition at the solid-liquid interface \cite{Navier,Bocquet2007}
\begin{equation}
v_t=b  \frac{\partial v_t}{\partial z},
\label{Navier}
\end{equation}
which relates  the tangential velocity of the fluid relative to the solid; $v_t$,  to the shear rate  at the wall; $b$ is the so-called slip length, and $z$ is the coordinate along the normal to the wall.

The slip length can be directly interpreted in terms of the liquid-solid friction at the interface. Indeed, in the presence of boundary slippage, the friction force at the liquid-solid interface is expected to be linear in slip velocity $v_t$:
\begin{equation}
F_w= - {\cal A}\, \lambda \, v_t
\label{friction}
\end{equation}
with  $\lambda$ the liquid-solid friction coefficient and ${\cal A}$ the lateral area. The Navier boundary condition can be interpreted as a balance between the viscous stress in the fluid, behaving as $\eta {\partial_z v_t}$ (with $\eta$ the bulk viscosity) and the solid-liquid friction force at the wall, $\lambda \, v_t$ (per unit surface). Accordingly  the slip length is related to the
friction coefficient according to $b=\eta/\lambda$.

As any phenomenological transport coefficient that describes linear response,  $\lambda$ is expected to be written in terms of {\it equilibrium properties}, in the form of a Green-Kubo (GK) relationship. Previous derivations of such GK relations for the slip length were developed on the basis of linear response applied to non-equilibrium shear drive described by a perturbed Hamiltonian  on one hand, and on the basis of a projection formalism on the other hand 
\cite{BarratBocquetPRE94}. Both derivations lead to the following expression for the friction coefficient in terms of the surface lateral force autocorrelation function (ACF) at equilibrium:
\begin{equation}
\lambda= {1\over {\cal A}\, k_BT} \int_0^\infty dt\, \langle F_w(t)\cdot F_w(0)\rangle_{\rm equ}
\label{GK}
\end{equation}
where $F_w$ is the total (microscopic) lateral force acting on the wall surface.

Note that the applicability of such formula raises some delicate and subtle questions, as was emphasized by Petravic and Harrowell \cite{Harrowell}. This is due to the order in which limits are taken between system size (number of fluid particles) and time going to infinity, as previously demonstrated by Espa\~nol and Zu\~niga \cite{Espanol} for diffusion 
and by Bocquet {\it et al.} for colloidal friction for several particles \cite{Bocquet1997}. Spurious results may be obtained from the calculations of GK integrals if the order of limits is not taken properly, see {\it e.g.} \cite{Bocquet1997}.

The GK relation nevertheless  provides a very useful framework in order to extract the dependence of the slip length from the microscopic properties, such as  the interfacial structure, density and interaction of the liquid at the interface \cite{Bocquet2007}, as well as some more subtle parameters such as commensurability and curvature \cite{Falk, Falk12}. This can be achieved using either approximate calculations, or a numerical integration using trajectories generated in molecular dynamics simulations, withe the usual difficulties associated with integrating time correlation functions determined in finite systems. 

In this note, we propose a new, alternative derivation for the GK relation of the friction coefficient, Eq.(\ref{GK}). This derivation is based on a Langevin approach for the Brownian motion of a massive wall in contact with a fluid. It points to the subtle non-markovian effects associated with the relaxation of hydrodynamic modes in the fluid.

\section{Non-markovian Langevin equation}

The geometry of the system is the following. We consider a (planar or cylindrical) wall in contact with a fluid, and assign a (large) mass $M$ to the wall. % in the end will be set to infinity. 
%with shear viscosity $\eta$ (kinematic viscosity $\nu$). 
The wall is assumed to be invariant under translation in at least in one direction. %along which the system is . 

In the presence of liquid-solid friction, the Langevin equation for the fluctuating wall velocity $U(t)$ takes the form:
\begin{equation}
M{d U\over dt}= - \lambda {\cal A} v_s(t) + \delta F(t)
\end{equation}
with {\cal A} the lateral surface, $\delta F(t)$ the lateral fluctuating force, $\lambda$ the wall-fluid friction
coefficient; $v_s(t)$ is the hydrodynamic slip velocity at the surface, defined here as
the difference between the wall and fluid velocity $v_s(t)=U(t)-v_f(t)$.

In the linear response regime,  the slip velocity is linearly related to the velocity $U(t)$ of the wall. with a general relation in the form
\begin{equation}
v_s(t)=\int_{-\infty}^{+\infty} dt'\, \xi(t-t') U(t')
\label{defxi}
\end{equation}
$\xi(t)$ is a memory kernel, which takes its origin in the relaxation of hydrodynamic (shear) modes in the fluid. Specific examples in the  case where inertial effects are neglected, and the fluid can be described by the Stokes equation, will be discussed in section \ref{examples}
Back into the Langevin equation, one obtains 
\begin{equation}
M{d U\over dt}= - \lambda {\cal A} \int_{-\infty}^\infty dt'\, \xi(t-t') U(t')
+ \delta F(t)
\label{Langevin}
\end{equation}

The memory function $\xi(t)$ should  obey several general relationships, which will prove useful in the derivation of a Green Kubo relation. 
A first, obvious property, is causality, which imposes that
$\xi(t)=0$ for $t<0$ so that the upper limit of the  integral in equations (\ref{defxi}) and (\ref{Langevin}) is $t$. A second property stems from the fact that the response of the fluid cannot be instantaneous.
Consider a situation at rest and a step change in wall velocity from $0$ to $U$ at time $t=0$. The fluid velocity cannot 
follow this step variation instantaneously and $v_f(t=0^+)=0$. Accordingly, $v_s(0^+)=U(0^+)-v_f(0^+)=U(0^+)$.
This shows that the Fourier (and Laplace) transform of $\xi(t)$ obeys the sum rule $\tilde\xi(\omega \rightarrow \infty)=1$ (or equivalently that the kerne contains a $delta(t-t')$ contribution). 
Another sum rule originates from galilean invariance. The Langevin equation, Eq.(\ref{Langevin}), should not be changed
by shifting the wall velocity by a constant $U_0$. This imposes that the time integral of $\xi$ vanishes:
$\tilde\xi(\omega=0)=\int_{-\infty}^\infty \xi(t)\,dt=0$.
All these properties will be specifically verified in the calculations of section \ref{examples}, based on the Stokes dynamics for the fluid.

In addition, the Fluctuation-Dissipation theorem (FDT) imposes a relation between the memory function $\xi(t)$ and the autocorrelation of the random force, 
\begin{equation}
\langle \delta F(t)\delta F(0) \rangle = 2\lambda {\cal A} k_BT \xi(t) \ \ \textrm{ for} \ \ t>0.
\label{FDT}
\end{equation}
with a vanishing integral  \footnote{As a side remark, note also that the calculation implicitly assumes a system with an infinite number of particles $N=\infty$, so that the thermodynamic
limit is already achieved, before any time or mass limit are taken.}.

%The non-markovian Langevin equation for the wall velocity takes the final form 
%\begin{equation}
%M{d U\over dt}= - \lambda {\cal A} \int_{-\infty}^t ds\, \xi(t-s) U(s)
%+ \delta F(t)
%\end{equation}

\section{Green-Kubo expression for the friction coefficient}

%
%Back into the Langevin equation, one obtains 
%\begin{equation}
%M{d U\over dt}= - \lambda S \int_{-\infty}^t ds\, \xi(t-s) U(s)
%+ \delta F(t)
%\end{equation}

%Note that the Fluctuation-Dissipation theorem imposes
%\begin{equation}
%\langle \delta F(t)\delta F(0) \rangle = 2\lambda S k_BT \xi(t)
%\end{equation}
%with a vanishing integral.

We consider the general situation, whereby the wall velocity $U(t)$ obeys the non-markovian Langevin equation,
Eq.(\ref{Langevin}), with the FDT in Eq.(\ref{FDT}), and complemented by the sum rules introduced above.

The Langevin equation is best analyzed by Laplace transform. Note that the Laplace transform is defined here for any complex number $s$ with $Re(s)> 0$ as
\begin{equation}
\tilde C(s)=\int_{0^-}^{+\infty} dt\, f(t)\, \exp[-s t]
\end{equation}
with the lower bound as $t=0^-$ (instead of $t=0^+$). This is crucial to account for the short time behavior \footnote{Calculation with the classical Langevin equation, for which  $\xi(\omega)=\xi_0$, do highlight this subtlety.}.

The Laplace transform of the velocity autocorrelation function $C(t)=\langle U(t)U(0)\rangle$ takes the expression 
\cite{BarratHansen}:
\begin{equation}
\tilde C(s)= {k_BT\over M} \left(s+ \alpha  \tilde\xi(s) \right)^{-1}
\end{equation}
where $\alpha= \lambda {\cal A}/M$ has the dimension of an inverse time scale, and we used $C(t=0)=k_BT/M$ from equipartition.

%{\color{red} Problem:  $\int_{-\infty}^{+\infty} \xi(t) dt=\xi(\omega=0)=0$
%Thus the Force-Force autocorrelation function is vanishing !! But not necessarily $\lambda$
%the friction coefficient...}

%A tester:
%two walls with two different velocities $U_1(t)$ and $U_2(t)$ ? Cf Harrowell ??
%\begin{equation}
%\lambda= {1 \over 2Sk_BT} \int_{-\infty}^{+\infty} \langle \delta F(t)\delta F(0) \rangle dt
%\end{equation}

Our aim is now to compute the time integral of the force autocorrelation function,
$\int_0^\infty \langle F_w(t)\cdot F_w(0)\rangle$ (with $F_w$ the force acting on the wall along a translation invariant direction,  
 say $x$), and show its relation to the friction coefficient $\lambda$.

The force acting on the wall along the direction $x$ is  $F_w(t)= M{dU\over dt}$ and its autocorrelation function  can be written as 
\begin{equation}
\gamma(t)=\langle F_w(t)\cdot F_w(0) \rangle= -M^2 {d^2 \over dt^2} \langle  U(t)\cdot U(0)\rangle. 
\end{equation}
Its Laplace transform, $\tilde\gamma(s)$
is then easily calculated from the expression of $\tilde C(s)$ as:
\begin{equation}
\tilde\gamma(s)=-M^2\left\{ s^2 \tilde C(s) - sC(t=0^-) -C^\prime (t=0^-)\right\}
\label{gammaz}
\end{equation}

The various terms in  equation  \ref{gammaz} can be calculated separately. First, $C(t=0)=k_BT/M$ from equipartition as above.
 Second, $C^\prime (t=0^-)=-C^\prime (t=0^+)=-\lim_{s\rightarrow \infty}s\times [s\tilde C(s)-C(t=0)]$. This can be rewritten as
\begin{equation}
C^\prime (t=0^-)= \alpha {k_BT\over M} \lim_{s\rightarrow \infty} {s\times \tilde{ \xi}(s)\over s + \alpha \tilde{\xi}(s)}
\end{equation}

Using the previously introduced  sum rule giving $\xi(s\rightarrow \infty)=1$, one deduces $C^\prime (t=0^-)= \alpha\times {k_BT/ M}$.
Altogether this leads to the following expression for the Laplace transform of the force autocorrelation function:
\begin{equation}
\tilde\gamma(s)=M \alpha  k_BT \left\{ 1+{s\tilde\xi(s)\over {s+\alpha \tilde\xi(s)}}\right\}.
\end{equation}

Taking the $s\rightarrow 0$ limit, we obtain finally:
$\tilde\gamma(z=0)= k_BT \lambda S$, {\it i.e.}
\begin{equation}
\lambda = {1\over {\cal A} k_BT} \int_0^{\infty}dt\, \langle F_w(t)\cdot F_w(0) \rangle
\end{equation}

This is just the previously obtained Green-Kubo relationship for the friction coefficient \cite{BarratBocquetPRE94}. The above derivation extends to a general friction kernel the proof given in Ref.  \cite{barratchiaruttini} for the closely related problem of the Kapitza resistance at a liquid solid interface, which was restricted to 
 a purely Markovian description.

\section{Examples: Stokes dynamics for planar and cylindrical geometries}
\label{examples}

It is interesting to compute explicitly the friction kernel $\xi(t)$ for specific cases, assuming a simple description of the dynamics in the fluid. 
In particular, we will consider a fluid with dynamics described   by the Stokes equation 
velocity $v_f(t)$.
\begin{equation}
\rho \partial_t \vec{v}_f(\vec{r},t) = -\nabla P + \eta \Delta \vec{v}_f(\vec{r},t)
\end{equation}
where $\eta$ is the shear viscosity and $\rho$ the mass density. The kinematic viscosity is defined as $\nu=\eta/\rho$. 
This equation is complemented by the Navier boundary condition at the wall surface, Eq.(\ref{Navier}). Although this description is not a microscopic one,
 it is known to describe correctly the fluctuations in any simple fluid in the hydrodynamic (long time, long wavelength) limit. As such, we expect that the properties of $\xi(t)$ derived using this simple model are general in the long time limit, while the short time behaviour will obviously depend on the specific wall-fluid interactions.

In the following, we  assume that the wall velocity exhibits a a time-dependent sinusoidal variation
along a given direction, say $x$, $U(t)=U_\omega e^{j \omega t}$. The kernel $\xi$ is obtained by analyzing the motion of the fluid in response to this imposed sinusoidal motion.

\subsection{Planar geometry}
In a planar geometry for the wall, the velocity profile in the fluid is deduced as:
\begin{equation}
v_\omega(z)={U_\omega \over {1 + b/\delta_\omega}} e^{-z/\delta_\omega}
\label{vomega}
\end{equation}
with $z$ the direction perpendicular to the wall, $b$ is the slip length and $ \delta_\omega^{-1}=\sqrt{j\omega \rho/\eta}$ is the size of the viscous boundary layer.

We thus deduce the slip velocity $v_s(t)=U(t)-v(z=0,t)$ as 
\begin{equation}
v_s(t)=\int {d\omega \over 2\pi} e^{-j\omega t} \left(1-{1\over 1 + b\sqrt{j\omega/\nu}} \right) U(\omega)
\end{equation}
which can be rewritten as a convolution of the wall velocity $U(t)$:
\begin{equation}
v_s(t)=\int_{-\infty}^{+\infty} ds\, \xi(t-s) U(s)
\end{equation}
with $\xi(t)$ the inverse Fourier transform of 
\begin{equation}
\tilde\xi(\omega)=1-{1\over 1 + b\sqrt{j\omega/\nu}}.
\end{equation}
Note that the Laplace transform of $\xi(t)$ is $\tilde\xi(s)=b\sqrt{s/\nu}/(1+b\sqrt{s/\nu})$.

One may verify that the inverse Fourier Transform of the above expression for $\tilde\xi(\omega)$  is given by
\begin{equation}
\xi(t)=\delta(t)-\theta(t) \left( {1\over \sqrt{\pi} \sqrt{\nu t/b^2}} - e^{\nu t/b^2} {\rm Erfc}[\sqrt{\nu t/b^2}] \right)
\end{equation}
with $\delta(t)$ the Dirac distribution and $\theta(t)$ the Heaviside function; ${\rm Erfc}$ is the complementary error function $ {\rm Erfc}(x)=1- {\rm Erf}(x)$ and
$ {\rm Erf}(x)=2/\sqrt{\pi} \int_0^x dt\, \exp(-t^2)$.
%\footnote{This expression can be found by remarking that the inverse Fourier Transform of $\xi(\omega)$
%can be transformed in terms of an inverse Laplace transform using $s=i\omega$. Reversly one may
%indeed check that the Fourier transform of the above expression for $\xi(t)$ is indeed $\xi(\omega)$...}

Note that %causality is indeed preserved, as it should: $\xi(t)=0$ for $t<0$ ({\it cf.} the Heaviside function).
all properties announced above for $\xi(t)$ are indeed verified by the above expression: causality, as well as 
the two sum rules $\tilde\xi(\omega=0)=0$, $\tilde\xi(\omega=\infty)=1$.

As a side remark, the memory kernel can be shown from the previous expression to decay algebraically for long times, as $\xi(t)\sim - {1\over{2 \sqrt{\pi}}}\left({\nu t\over b^2}\right)^{-3/2}$. While algebraic decays are expected for the relaxation of hydrodynamic modes, the exponent, $3/2$, is not usual for a (basically) 1D diffusion problem.

Another intriguing consequence is that, according to the results of Morgado {\it et al.} \cite{Hansen2002}, the fluctuating wall position, $X(t)$ (with $\dot X(t)=U(t)$), is expected to undergo superdiffusive motion as 
\begin{equation}
\langle X^2(t) \rangle \sim t^{3/2}
\end{equation}
and the diffusion coefficient is accordingly infinite.
This is an immediate consequence of the vanishing integral of the friction coefficient with $\tilde\xi(\omega)\sim \omega^{\beta-1}$, and $\beta >1$ \cite{Hansen2002}. In the present geometry, $\tilde\xi(\omega)\sim \omega^{1/2}$ and $\beta=3/2$.  This anomalous behaviour can also be understood from the usual 'long time tail' arguments \cite{BarratHansen}: after a time $t$, the initial momentum of the wall is spread over a fluid thickness that scales as $t^{1/2}$, which leads to expect a velocity autocorrelation that scales as $t^{-1/2}$.

%in the present geometry.

\subsection{Cylindrical geometry}

If we now consider a cylinder geometry with radius $R$, the solution to the velocity profile  inside the cylinder 
takes the following expression,
\begin{equation}
v_\omega(r)={U_\omega \over {I_0(R/\delta_\omega) + b/\delta_\omega} I_1(R/\delta_\omega)} I_0(r/\delta_\omega)
\end{equation}
where $r$ is the  distance to the center, and the velocity is parallel to the axis.
Accordingly, the Fourier transform of the friction coefficient takes the form
\begin{equation}
\tilde\xi(\omega)=1-\left(1 + {b\over\delta_\omega} {I_1(R/\delta_\omega)\over I_0(R/\delta_\omega)}\right)^{-1}
\end{equation}
We could not find an explicit expression for the kernel in real time, $\xi(t)$. 
However, as for the planar geometry, one may verify explicitly that the kernel indeed obeys the expected sum rules $\tilde\xi(\omega=0)=0$, $\tilde\xi(\omega=\infty)=1$.

An interesting result for the cylindrical geometry concerns the low frequency behavior.
In this limit $\xi(\omega)\sim {b\cdot R / 2\delta_\omega^2}$, so 
that $\xi(\omega) \propto \omega$. Interestingly, this is different from the planar wall, 
for which $\xi(\omega)\propto \sqrt{\omega}$. Accordingly, in the cylindrical case, 
one expects a stronger time decay, as $\xi(t) \sim R\cdot b/(\nu t)^2$.

As for the planar geometry, the fluctuating wall position is superdiffusive, however with a different (larger) exponent
\begin{equation}
\langle X^2(t) \rangle \sim t^{2}
\end{equation}
Again, this is consistent with the usual long time tail arguments: the initial momentum of the wall is now spread over a finite volume, so that the velocity autocorrelation function tends to a constant in the long time limit. 
%Now, using the above relationship between $\tilde\gamma(z)$ and $\tilde\xi(z)$ (Laplace transform), one
%is left to the same conclusions as before: taking first the limit $\alpha\rightarrow 0$ for a fixed $z$ -- so that the
%infinite mass limit is taken before the infinite time limit --, then $\tilde\gamma(z)=\lambda S k_BT (1 - \tilde\xi(z))$
%for $\alpha\rightarrow 0$ and using $\tilde \xi(z=0)=0$, one deduces again 
%$\tilde\gamma_{M=\infty}(z=0)= k_BT \lambda S$, {\it i.e.}
%\begin{equation}
%\lambda = {1\over S k_BT} \int_0^{\infty}\langle F_w(t) F_w(0) \rangle
%\end{equation}
%which is also valid for a circular geometry.
\section{Conclusion}

In this brief paper, we have provided an alternative derivation for the Green-Kubo relationship for the solid-liquid
friction coefficient, based on a Langevin approach for the dynamics of a fluctuating wall. The derivation is based on
general arguments and do not suffer from some of the approximations involved in the previous derivations \cite{BarratBocquetPRE94}.
This reinforces accordingly the foundations of the framework describing solid-liquid friction, based on the derived Green-Kubo relation. The latter is at the root of various recent developments for fluid-solid incommensurability, leading to huge slippage effects for the water-carbon nanotube interface \cite{Holt06,Falk,Falk12}. 

The derivation also highlights some subtleties involved in the relaxation process for the fluid-wall interface, yielding algebraically decaying memory terms, with vanishing time integral. This leads to superdiffusive motion of the fluctuating wall position, with an exponent that depends on the geometry.

\end{document}